\documentclass[10pt]{article}
\usepackage[T1]{fontenc}

\usepackage{graphicx}

\usepackage{wrapfig}
\usepackage[T1]{fontenc} 

\usepackage[normalem]{ulem} 

\usepackage[T1]{fontenc} 
\usepackage[normalem]{ulem} 

\usepackage[usenames,dvipsnames,svgnames,x11names]{xcolor}
\usepackage{fontawesome5}

\usepackage[usenames,dvipsnames,svgnames,x11names]{xcolor}
\usepackage{fontawesome5}

\newcommand{\para}[1]{\textbf{#1.}~}

\newcommand{\method}[0]{\textsc{FedJoule}\xspace}

\newcommand{\random}[0]{\textbf{RND}\xspace}
\newcommand{\escs}[0]{\textbf{ESCS}\xspace}
\newcommand{\es}[0]{\textbf{ExSH}\xspace}
\newcommand{\ks}[0]{\textbf{KSH}\xspace}
\newcommand{\ilpes}[0]{\textbf{FedJ-ex}\xspace}
\newcommand{\ilpks}[0]{\textbf{FedJ-k}\xspace}

\usepackage{balance}

\usepackage{listings}

\lstdefinelanguage{XML}
{
basicstyle=\ttfamily\footnotesize,
  morestring=[b]",
  moredelim=[s][\bfseries\color{Maroon}]{<}{\ },
  moredelim=[s][\bfseries\color{Maroon}]{</}{>},
  moredelim=[l][\bfseries\color{Maroon}]{/>},
  moredelim=[l][\bfseries\color{Maroon}]{>},
  morecomment=[s]{<?}{?>},
  morecomment=[s]{<!--}{-->},
  commentstyle=\color{gray},
  stringstyle=\color{blue},
  identifierstyle=\color{red}
}

\lstdefinestyle{yaml}{
     basicstyle=\color{blue}\small\tt,
     rulecolor=\color{black},
     string=[s]{'}{'},
     stringstyle=\color{blue},
     comment=[l]{:},
     commentstyle=\color{black},
     morecomment=[l]{-}
 }
\usepackage{moreverb}

\usepackage[nounderscore]{syntax}

\usepackage[cmex10]{amsmath}
\usepackage{amssymb}
\usepackage{mathtools}
\usepackage{amsfonts}
\usepackage{gensymb}

\usepackage{subfig}
\usepackage{algorithmicx}
\usepackage{algpseudocode}
\usepackage[ruled]{algorithm}
\definecolor{light-gray}{gray}{0.75}
\algrenewcommand{\algorithmiccomment}[1]{\hskip3em{{\footnotesize \textcolor{light-gray}{$\blacktriangleright$}}} #1}

\usepackage{multirow} 
\usepackage{rotating} 
\usepackage{booktabs} 
\usepackage{colortbl} 
\usepackage{tablefootnote} 
\usepackage{longtable}
\usepackage{tabularx}

\usepackage{array}
\newcolumntype{L}[1]{>{\raggedright\let\newline\\\arraybackslash\hspace{0pt}}m{#1}}
\newcolumntype{C}[1]{>{\centering\let\newline\\\arraybackslash\hspace{0pt}}m{#1}}
\newcolumntype{R}[1]{>{\raggedleft\let\newline\\\arraybackslash\hspace{0pt}}m{#1}}

\usepackage[pdftex,colorlinks=true,urlcolor=blue,citecolor=blue]{hyperref}

\usepackage{xspace}

\usepackage{enumitem}

\usepackage{blindtext}

\begin{document}
\title{Federated Learning within Global Energy Budget over Heterogeneous Edge Accelerators\thanks{~Preprint of paper to appear in the proceedings of the 31st International European Conference on Parallel and Distributed Computing (EuroPar) 2025: Roopkatha Banerjee, Tejus Chandrashekar, Ananth Eswar and Yogesh Simmhan, “Federated Learning within Global Energy Budget over Heterogeneous Edge Accelerators,” in \textit{31st International European Conference on Parallel and Distributed Computing (EuroPar)}, 2025}}

\author{Roopkatha Banerjee, 
Tejus Chandrashekar, 
Ananth Eswar \\ 
and Yogesh Simmhan\\
Department of Computational and Data Sciences (CDS),\\
Indian Institute of Science (IISc),\\
Bangalore 560012 India\\
Email: \{roopkathab, simmhan\}@iisc.ac.in
}

\date{}
\maketitle       

\begin{abstract}
Federated Learning (FL) enables collaborative model training across distributed clients while preserving data privacy. However, optimizing both energy efficiency and model accuracy remains a challenge, given device and data heterogeneity. Further, sustainable AI through a global energy budget for FL has not been explored. We propose a novel optimization problem for client selection in FL that maximizes the model accuracy within an overall energy limit and reduces training time. We solve this with a unique bi-level ILP formulation that leverages approximate Shapley values and energy--time prediction models to efficiently solve this. Our \method framework achieves superior training accuracies compared to SOTA and simple baselines for diverse energy budgets, non-IID distributions, and realistic experiment configurations, performing 15\% and 48\% better on accuracy and time, respectively. The results highlight the effectiveness of our method in achieving a viable trade-off between energy usage and performance in FL environments.
\end{abstract}

\setlength{\belowdisplayskip}{0pt} \setlength{\belowdisplayshortskip}{0pt}
\setlength{\abovedisplayskip}{0pt} \setlength{\abovedisplayshortskip}{0pt}


\section{Introduction}

\textit{Federated Learning~(FL)} has transformed distributed Machine Learning (ML) by enabling privacy-preserving model training across decentralized edge devices, without moving data centrally. In FL, clients train local models using their private data and only share model updates with a central server for aggregation into a global model (Fig.~\ref{fig:fl-lifecycle}). This repeats over multiple rounds, using different subsets of clients, till convergence. FL is proving essential, given the pervasive generation of data from diverse and accelerated edge devices and the growing regulatory requirements like GDPR on data privacy.

\para{Context}
However, FL poses several challenges due to the \textit{heterogeneity of edge devices and of data} present across them, which affects the training time, energy usage and accuracy achieved.
Edge devices ranging from Raspberry Pi to GPU-accelerated NVIDIA Jetsons differ by orders of magnitude in compute performance and peak power load, which affects time to accuracy~\cite{sys-diversity}.
Further, Jetsons offer 1000s of \textit{power modes}, which control the active CPU cores and CPU/GPU/memory frequencies, with varying compute--power trade-offs that impact the energy footprint for FL~\cite{sigmetrics-prashanthi}. Non-IID data with label and quantity skews across edge devices can cause model drift, increasing training rounds~\cite{data-diversity}. 

\textit{Time to accuracy (TTA)}, the time required to achieve a target accuracy over 100s of rounds of training, is a critical metric for FL. Simultaneously, the focus on \textit{sustainable AI} has highlighted the need for energy-efficient FL. Enterprise-wise carbon caps and IoT field deployments, possibly powered by renewables, may impose collective energy budgets on FL training. Unlike energy constraints imposed on individual devices due to battery limits or power load, \textit{\textcolor{blue!80}{imposing a collective energy constraint for sustainable FL training is a novel problem that has not yet been sufficiently explored}}.

\para{Challenges}
NVIDIA Jetson accelerated edge devices are well-suited for FL. Besides multi-core Arm CPUs, they have a GPU with 100-1000s of CUDA cores and 8--32GB of RAM shared by CPU and GPU (Table~\ref{tbl:jetson_device_specs}), allowing them to train non-trivial DNN models with FL. Besides a compact size and a peak power within 60W, they also expose \textit{custom power modes} to change the CPU core count and frequencies of CPU, GPU and memory.
E.g., the latest Jetson Orin AGX has $29$ CPU frequencies, $13$ GPU frequencies, $4$ memory frequencies and $12$ core-counts to give $\approx 18,000$ power modes, with \textit{MAXN} being the highest power and performance one. While this can help achieve compute--energy trade-offs, the power mode settings have a non-linear impact on training time and power load, making it challenging to effectively tune this. Profiling of the power modes is time-consuming, taking minutes per power mode for each DNN being trained.
Hence, it is challenging to select the right subset of edge accelerated clients and their relevant power modes in each FL round to meet the energy budget and minimize TTA. This is further exacerbated by the non-IID nature of data distribution across clients that affects convergence.

\para{Contributions} 
We address these challenges through a novel framework, \textit{\method},and make the following contributions:

\begin{enumerate}[leftmargin=*]
\item We formulate a novel \textit{optimization problem} to select clients and their power modes in each FL round from a set of heterogeneous clients with non-IID data distribution such that the model accuracy is maximized while staying within a \textit{global energy budget} and we minimize the training time (\S~\ref{sec:problem}). 
\item We solve this optimization problem tractably by decomposing it into two parts solved using Integer Linear Programming (ILP)(\S\ref{sec:method:ilp}):\textit{(i)ILP-CS:} In each round, we select a subset of clients to maximize improvement in global model accuracy using approximate Kernel-based Shapley values per device while staying within the peak energy budget at MAXN mode (\S~\ref{sec:method:shapley}), and introduce a \textit{cooldown period} for the selected clients.\textit{(ii)ILP-PM:} We use our prior work, PowerTrain, to build a valuable time--energy Pareto front across power modes (\S~\ref{sec:method:powertrain}) to help pick the best power mode for the FL clients.
\item We evaluate the proposed solution using real-world traces from 4 Jetson device types for three popular DNN models (LSTM, MobileNet, ResNet), and simulate FL training on a heterogeneous cluster of 12 and 48 edge devices (\S~\ref{sec:results}). 
Our comparison against simple and state-of-the-art~(SOTA) baselines, FedAvg~\cite{McMahan2016CommunicationEfficientLO} and ESCS~\cite{escs}, shows \method is 48\% faster and reaches a 15\% higher accuracy for the given energy budget, enabling sustainable AI. 
\end{enumerate}
Besides these, we discuss related work in \S~\ref{sec:related} and offer our conclusions in \S~\ref{sec:conclude}.


\section{Related Work}

\label{sec:related}

\begin{wrapfigure}{r}{0.53\textwidth}

    \centering
    \includegraphics[trim={4cm 4.5cm 3cm 3.5cm},width=1\linewidth]{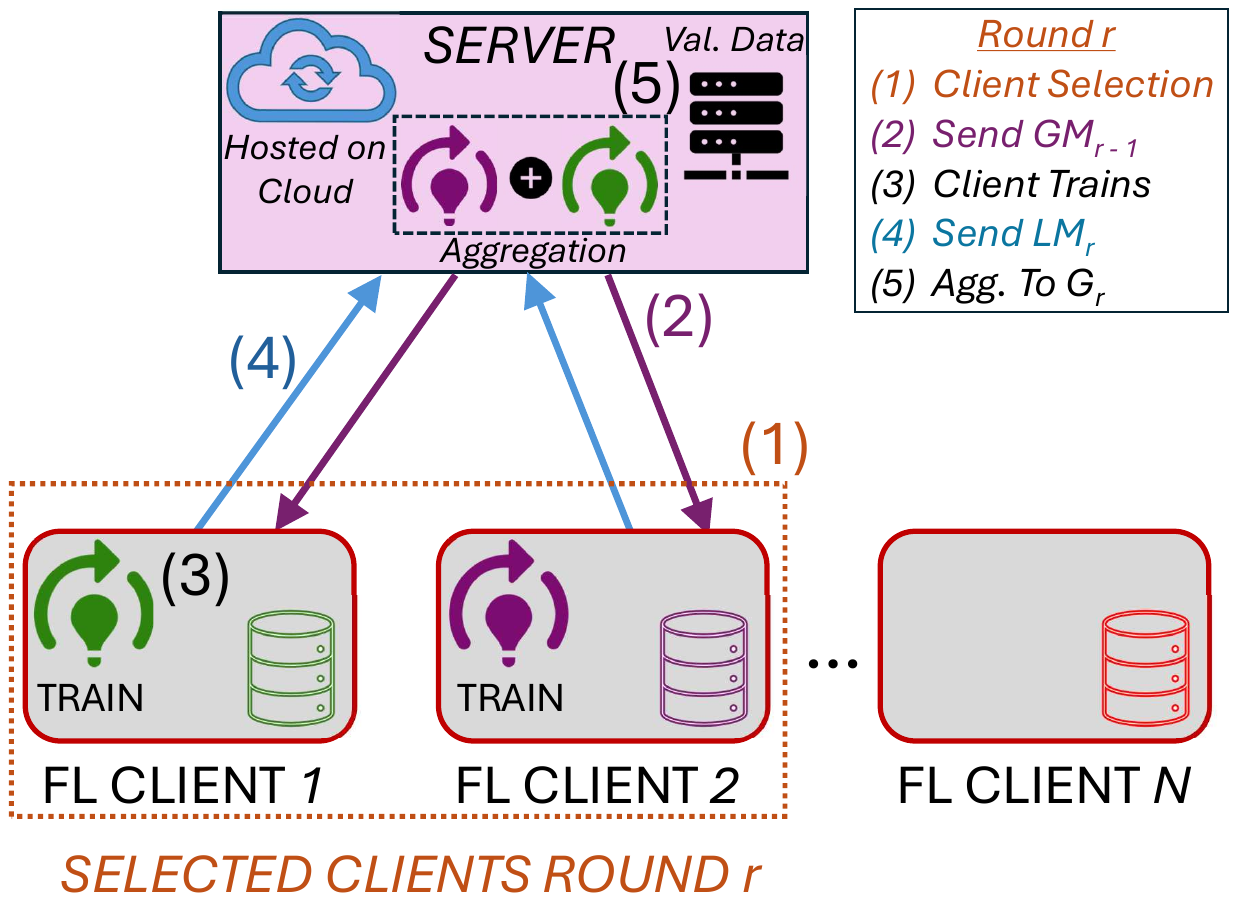}%
    
    \caption{Typical FL Lifecycle.}
    \label{fig:fl-lifecycle}
    
\end{wrapfigure}

Fig.~\ref{fig:fl-lifecycle} shows a typical FL lifecycle. We use FedAvg~\cite{McMahan2016CommunicationEfficientLO}, which is simple and effective under varied conditions compared to more complex strategies. It uses synchronous aggregation, where all clients report their local models before the global model is created from their weighted average. Several variations have been proposed to improve convergence with \textbf{data heterogeneity}~\cite{fedprox}.
However, such methods often lead to biased model updates. 
Importance sampling is used to prioritize clients with under-represented or statistically significant data to reduce bias in the aggregated model~\cite{cho2022towards}. Also, strategies to mitigate system heterogeneity, such as adaptive client participation, model compression, and asynchronous training. But most FL strategies employ client selection methods that remain inherently biased. In our experiments, \method out-performs both an unbiased approach, FedAvg~\cite{McMahan2016CommunicationEfficientLO}, which results in a generalized model but with poor convergence, and a biased one, ESCS~\cite{escs}, which converges faster but generalizes poorly.

Game-theoretic approaches such as \textbf{Shapley values}~\cite{wang2020principled} have emerged as a promising method to quantify marginal client contributions in FL.  
However, exact Shapley computation has exponential time complexity in the number of clients, making it intractable at scale. Some have proposed approximations such as truncated Monte Carlo sampling~\cite{jia2019towards} and gradient-based approximations~\cite{zhang2021gtg}.
We overcome this limitation in three ways: 
we use historical \textit{surrogate Shapley values} to avoid full client participation~\cite{sum_2023_shapleyFL}, we use kernel-based \textit{approximate Shapley}~\cite{lundberg_2017_kernelshap} to lower time taken while retaining client fairness, and use \textit{Coreset sampling} of the validation data used in Shapley calculations.

Recent research in \textbf{energy-constrained FL} focuses on limiting per-device energy or power load during training. Energy optimization is achieved via client selection strategies~\cite{escs} and resource optimization~\cite{chen2023eefl}. 
Some~\cite{liu2023green} reduce energy using hierarchical clustering that avoids redundant compute and communication. However, most methods use an idealized theoretical estimate of energy usage by the devices, 
and ignore power modes exposed by edge accelerators. Lastly, there is limited research on constraining the total energy across all FL training rounds. Yu et al.~\cite{Yu2022energyaware} propose a global energy-aware device scheduler that optimizes client participation and communication. While most prior works on leveraging Jetson power modes are limited to DNN inferencing, Prashanthi et al.~\cite{prashanthi2024powertrain} use transfer learning to estimate the latency and power usage of Jetson power modes for training, which we leverage in this paper.


\section{Problem Formulation}
\label{sec:problem}

\paragraph{Preliminaries.}
We consider a standard FL setting where a single leader coordinates the training process across a set $\mathcal{C}=\{c_1, c_2, \ldots, c_n\}$ of $n$ \textit{clients}. Each client $c_i$ possesses a local dataset $\mathcal{D}_i = {(x_{i,j}, y_{i,j})}, {j=\{1,\hdots,|\mathcal{D}_i|\}}$, where $x_{i,j}$ represents the features and $y_{i,j}$ the corresponding label. The \textit{global objective} in FL is to minimize the empirical loss over all the clients' datasets, $\min_{\boldsymbol{w}} F(\boldsymbol{w}) = \sum_{i=1}^{n} \frac{|\mathcal{D}_i|}{|\mathcal{D}|} F_i(\boldsymbol{w}, \mathcal{D}_i)$, where $|\mathcal{D}| = \sum_{i=1}^{n} |\mathcal{D}_i|$ is the total number of samples, $\boldsymbol{w}$ represents the \textit{model parameters}, and $F_i(\boldsymbol{w})$ is the \textit{local objective function} for client $i$: $F_i(\boldsymbol{w}, \mathcal{D}_i) = \frac{1}{|\mathcal{D}i|} \sum_{(x,y) \in \mathcal{D}_i} \ell(\boldsymbol{w}; x, y)$, with $\ell(\boldsymbol{w}; x, y)$ being the \textit{loss function} for a data point.

In each \textit{communication round} $r$, the leader \textit{selects a subset of clients} $\mathcal{S}^r \subseteq \mathcal{C}$ to participate in the training, limited to a \textit{maximum client count} of $\gamma$ per round. Each selected client $c_i \in \mathcal{S}^r$ receives the current global model $\boldsymbol{w}^r$ and performs $\eta$ local epochs of \textit{Stochastic Gradient Descent (SGD)} to obtain an \textit{updated local model} $\boldsymbol{w}_i^{r+1}$, returned to the leader. The leader aggregates all the local models to form the new global model $\boldsymbol{w}^{r+1}$ using a \textit{weighted average}, $\boldsymbol{w}^{r+1} = \sum_{c_i \in \mathcal{S}^r} \frac{|\mathcal{D}i|}{|\mathcal{D}_{\mathcal{S}^r}|} \boldsymbol{w}_i^{r+1}$, where $|\mathcal{D}_{\mathcal{S}^r}| = \sum_{c_i \in \mathcal{S}^r} |\mathcal{D}_i|$ is the total number of samples across all selected clients. Let the accuracy of the global model after round $r$ of training be $A^r$.

Let $\mathcal{P}_i$ be the set of all \textit{power modes} available for client $c_i$, and for
power mode $p_j \in \mathcal{P}_i$, let the corresponding \textit{time to train} one round of the above local model be $\tau_{ij}$ and $\epsilon_{ij}$ be its \textit{energy consumption}. Let $Z^r_{ij}$ be a binary decision variable that is $1$ when client $c_i \in \mathcal{S}^r$ selected in round $r$ operates using power mode $p_j$, and $0$ otherwise; only one power mode is selected, $\sum_{p_j\in\mathcal{P}_i} Z^r_{ij} = 1$. 
Hence, the \textit{local energy} consumed by the client $c_i$ for local model training in a round $r$ will be $E^r_i = \sum_{p_j \in \mathcal{P}_i} Z^r_{ij} \cdot \epsilon_{ij}$ and the \textit{energy for the round} across all selected clients is then $E^r = \sum_{c_i \in \mathcal{S}^r} E^r_i$. Here, we assume that the energy expended by the leader is negligible since it is idle for most of the round, except when selecting clients and aggregating.

The training time for the round is determined by the slowest of the clients, since typically, the aggregation time is much smaller and constant. The \textit{round training time} is $T^r = \max_{c_i \in \mathcal{S}^r} Z^r_{ij} \cdot \tau_{ij}$.


\paragraph{Optimization Problem.}
Given set of clients $\mathcal{C}$, the maximum clients per round $\gamma$ and an overall energy budget $\widehat{E}$ for the FL training, 

\begin{equation*}
\begin{array}{l}
\text{Select a set of clients in each round } r \text{ of FL training, } \mathcal{S}^r \in \mathcal{C} \text{ and } |\mathcal{S}^r| \leq \gamma \text{, and}\\
\text{Select the set of power modes for the clients, } Z^r_{ij} \text{, and}\\
\text{Find the maximum number of training rounds } r'\\
\text{s.t. the \textit{global energy consumed} for FL is within the budget, } \sum_{r \leq r'} E^r \leq \widehat{E} \text{, and}\\
\text{we \textit{maximize the accuracy achieved}, } \arg\max A^{r'}\text{ as a primary goal, and}\\
\text{we \textit{minimize the training time}, } \arg\min \sum_{r \leq r'} T^r \text{ as a secondary goal.}\\
\end{array}
\end{equation*}


\paragraph{Assumptions.}
We make certain simplifying and reasonable assumptions. We assume the system is not network-constrained. While prior works show wireless communication to account for a significant energy use in federated learning, we assume a stable high-speed LAN connection. We see that the communication energy is $\leq 0.2\%$ of the computation energy, based on the relative time spent in communication. Modelling communication energy is left to the future. Similarly, the aggregation server is assumed to run on the cloud, where energy is not a constraint and there is a push towards using sustainable power sources. The framework operates in a trusted environment with pre-deployed models and data. We use an honest-but-curious threat model where both the server and clients behave faithfully, and defer more adversarial settings to future work. We use a synchronous, non-hierarchical FL architecture with FedAvg for aggregation.


\section{Methodology}
\label{sec:method}

We decompose the optimization into two sub-problems: selecting clients per round~(\textit{ILP-CS}) and choosing their power modes (\S\ref{sec:method:ilp})~(\textit{ILP-PM}), both solved efficiently using Integer Linear Programming (ILP). To support this, we extend our prior work on PowerTrain~(\S\ref{sec:method:powertrain}) to generate a \textit{energy vs. time Pareto} for local training on each edge device type. During a bootstrap phase, we profile DNN training across tens of power modes per device and build a prediction model that estimates training time ($\tau_{ij}$) and energy consumption ($\epsilon_{ij}$). The resulting Pareto front provides \textit{candidate power modes} for the \textit{ILP-PM} solution.

For each FL round, we \textit{select clients} based on the improvements they offer to the global model quality while staying within overall energy budget. This choice is based on the \textit{Kernel Shapley} values for a client cohort (\S\ref{sec:method:shapley}) which address the high computational costs of (exhaustive) Shapley values. We further introduce a \textit{cool-down factor} that avoids selecting the same clients repeatedly.
These components are jointly formulated as \textit{ILP-CS} (Eqn.~\ref{eqn:ilp}).
For these \textit{select clients}, we optimize their \textit{power mode selection} based on the Pareto plot, 
formulated as \textit{ILP-PM} (Eqn.~\ref{eqn:opt}), that reduces the energy usage for the round. 
Once the round training is completed, we estimate the global energy budget that is still remaining, and if available, we start the next round of training. This repeats until we exhaust the energy budget. So the number of rounds is indirectly derived.


\subsection{Solving Bi-Level Optimization using ILP}
\label{sec:method:ilp}

We reformulate the problem as a bi-level optimization, significantly reducing the size of the solution space. The solution space decreased from 
worst case $\mathcal{O}\left((1+|\mathcal{P|})^{|\mathcal{C}|}\right)$, choosing $\gamma$ clients from $|\mathcal{C}|\simeq 100$ each with $|\mathcal{P}|\simeq 10,000$ power modes for the joint optimization problem, to one of 
$\mathcal{O}(2^{|\mathcal{C}|})$ for client selection alone followed by $\mathcal{O}(|\mathcal{P}'|^{|\mathcal{S}^r|})$ for power mode assignment alone, where $|\mathcal{P}'|\simeq 100$ are power modes on the energy-time Pareto.

We formulate the \textit{ILP-CS problem} to select clients $\mathcal{S}^r$ in round $r$ as:
\begin{equation}
\arg\min_{\mathcal{S}^r}~\big(\alpha T^r - (1-\alpha) A^r \big)
\quad \textrm{s.t.}\quad \dot{E}^r \leq \big(\widehat{E} - \sum_{i<r}E^i\big)
\label{eqn:ilp}
\end{equation}

Here, $E^i$ for $i<r$, is the energy used in prior rounds. For clients selected in the current round, we assume that they run on MAXN to compute the expected energy, $\dot{E}^r$. This formulation captures the trade-off between \textit{expected accuracy} of the global model, $A^r$, after the $r^{th}$ aggregation using local models from clients $\mathcal{S}^r$, and the \textit{expected time spent} in the round, $T^r$, with $\alpha\in[0.0,1.0]$ being a hyper\-parameter prioritizing speed over accuracy. 
Having a faster round time can also indirectly reduce the energy used in the round since it is a product of device time and device power.

We next formulate the \textit{ILP-PM problem} to decide the power modes, $Z^r_{ij}$, for the $\mathcal{S}^r$ clients selected above so that we reduce the expected energy further, but without increasing the round time.

\begin{equation}
\arg\min_{Z^r_{ij}} \sum_{c_i \in \mathcal{S}^r}\sum_{p_j \in \mathcal{P}_i}\big( Z^r_{ij} \cdot \epsilon_{ij} \big) 
\quad\textrm{s.t.}\quad
\sum_{p_j\in \mathcal{P}_i} Z^r_{ij} \cdot \tau_{ij} \leq T^r, \forall\text{i}\in\mathcal{S}^r\\
\label{eqn:opt}
\end{equation}

We pick a power mode lower than MAXN for a subset of clients to ensure that the \textit{actual energy used} is lower than $\dot{E}^r$. But we also avoid selecting too low a power mode to cause a client device to take longer to train its model than the slowest of the selected clients running at MAXN, thereby increasing the round time. Note that the choice of power modes will not affect the accuracy of the local model only its training time. Further, we introduce a custom \textit{cooldown factor} denoting the number of rounds a client will not be selected again once it has been selected in a round. The cooldown factor $\theta$ for a client $c_i\in\mathcal{S}^r$ is defined as $\theta_i = \lceil \rho\cdot A^r_i \rceil$, where $A^r_i$ is the training accuracy reported by the client's local model on its local data, and $\rho$ is a user-defined hyper\-parameter. 

As discussed earlier, our Pareto model (\S~\ref{sec:method:powertrain}) will give the energy and time estimates for a client's power mode.
We also need an estimate of $A^r$, the global model accuracy after a round for the selected clients. We propose to use the cumulative approximate Shapley values for the client cohort to estimate this (\S~\ref{sec:method:shapley}).
Given these, \method efficiently solves these two ILP problems at the start of each round using IBM ILOG CPLEX Optimizer to select the clients and their power modes for the round, while remaining within the energy budget.


\subsubsection{Exhaustive Shapley Values \textbf{(ExSH)}.} In our FL framework, we conceptualize each training round $r$ as a cooperative game $G(\boldsymbol{w}^r, \mathcal{C}, \mathcal{D}_v, \Psi, \mathcal{A})$, where $\boldsymbol{w}^r$ represents the global model parameters at the start of round $r$ and $\mathcal{C}$ are the set of available clients. The function $\Psi(\mathbf{w}^r, \mathcal{S}^r)$ returns the updated model parameters after aggregating contributions from client subset $\mathcal{S}^r$ using the update rule for $\boldsymbol{w}^{r+1}$ in \S~\ref{sec:problem}.
We define the \textit{utility of a cohort} $\mathcal{S}^r \subseteq \mathcal{C}$ as the performance of the updated global model on validation data, formalized as $A^r = \mathcal{A}(\mathcal{D}_v, \Psi(\mathbf{w}^r, S))$, where $\mathcal{A}$ evaluates the model's accuracy on dataset $\mathcal{D}_v$. The \textit{partial federated Shapley value} for client $c_i$ in round $r$ is then defined as:
\begin{equation}
\begin{aligned}
\phi_i^r 
= &\frac{1}{|\mathcal{S}|} \sum_{ s\subseteq \mathcal{S} \setminus \{c_i\}} \frac{\mathcal{A}(s \cup \{c_i\}) - \mathcal{A}(s)}{\binom{|\mathcal{S}|-1}{|s|}}  \\
= &\sum_{s \subset \mathcal{S} \setminus \{c_i\}} \frac{\mathcal{A}(\mathcal{D}_v, \Psi(\mathbf{w}^r, s \cup \{c_i\})) - \mathcal{A}(\mathcal{D}_v, \Psi(\mathbf{w}^r, s))}{|\mathcal{S}|\binom{|\mathcal{S}|-1}{|s|}}
\end{aligned}
\end{equation}
This quantifies the marginal contribution of $c_i$ to the model's accuracy across all possible client cohorts in $\mathcal{S}^r$, providing a principled measure of the client utility that considers both data quality and interaction effects among clients

However, as training progresses, the magnitude of performance improvements typically diminishes, creating unequal ranges of partial Shapley values across rounds. To address this, we borrow the \textit{normalization method} from \cite{sum_2023_shapleyFL} and apply min-max normalization within each round,
$\widehat{\phi}_i^r = \frac{\phi_i^r - \min(\boldsymbol{\phi}^r)}{\max(\boldsymbol{\phi}^r) - \min(\boldsymbol{\phi}^r)}$,
where $\boldsymbol{\phi}^r = \{\phi_i^r | c_i \in \mathcal{S}^r\}$ is the set of all partial federated Shapley values in round $r$, and $\max(\cdot)$ and $\min(\cdot)$ return the maximum and minimum values in the set, respectively. This normalization preserves the relative contributions of clients within each round, which is crucial for fair client selection.

We also maintain the client's historical contribution across rounds through a \textit{surrogate federated Shapley value}~\cite{sum_2023_shapleyFL}, as an exponential weighted average of past and current values,
$\tilde{\phi}_i^r = 
\begin{cases}
\beta \cdot \tilde{\phi}_i^{r-1} + (1-\beta) \cdot \widehat{\phi}_i^r, & \text{if } c_i \in \mathcal{S}^r \\
\tilde{\phi}_i^{r-1}, & \text{if } c_i \notin \mathcal{S}^r
\end{cases}$.
Here, $\beta \in [0.0,1.0]$ controls the update rate of the surrogate value, determining how much weight to place on past contributions by a client versus its recent performance. A smaller $\beta$ prioritizes recent contributions, allowing the system to adapt quickly to client changes, while a larger $\beta$ provides stable estimates over time.

Using Shapley values allows us to establish a fair mechanism for client selection. In our energy-aware optimization framework, Shapley values serve as the foundation for identifying high-value clients within the energy constraints of the system, allowing us to effectively select the subset $\mathcal{S}^r$ that maximizes the expected model accuracy $A^r$ while meeting our energy budget.


\subsubsection{Kernel-based Shapley Values for Efficient Computation \textbf{(KSH)}.}\label{sec:method:shapley}
Computing exact Shapley values require evaluating all possible client cohorts in $\mathcal{S}^r$, making it computationally prohibitive for more than a few clients. We use a fast kernel-based Shapley value approximation~\cite{lundberg_2017_kernelshap} that does importance sampling of the set of cohorts to be evaluated.
We assign weights to a cohort $S\subseteq \mathcal{S}^r$:
\begin{equation*}
w(|S|, |\mathcal{S}^r|) =
\begin{cases}
1, & \text{if }  |S| = |\mathcal{S}^r|, \\
\frac{|\mathcal{S}^r| - 1}{\binom{|\mathcal{S}^r|}{|S|} \cdot |S| \cdot (|\mathcal{S}^r| - |S|)}, & \text{otherwise}
\end{cases}
\end{equation*}
For our problem, we set the number of cohorts to be sampled as $\min(2^{|\mathcal{S}^r|},3\cdot|\mathcal{S}^r|)$. 
We use these weights to probabilistically sample $2\cdot|\mathcal{S}^r|$ 
unique cohorts to evaluate, which was found to be ideal through an ablation study. Further, we also evaluate each client model individually (singleton cohort), adding up to a total of $3\cdot|\mathcal{S}^r|$ evaluations every round. 

For the set of clients $\mathcal{S}^r$ that have trained in round $r$, let the set of selected cohorts be $\mathcal{X}^r$ and the evaluation accuracies of a cohort $S\in\mathcal{X}$ be $v_s=\mathcal{A} (\mathcal{D}_v,\Psi(w^r,S))$. Using these values, we then find the approximate Shapley values $\bar{\phi}_i$ for client $c_i\in\mathcal{S}^r$ by solving the weighted least-squares problem,
$\min_{\bar{\boldsymbol{\phi}^r}}  \sum_{s \in \mathcal{X}^r} w(|s|, |\mathcal{S}^r|) \cdot \left( v_s - \sum_{c_i \in \mathcal{S}^r} \phi^r_i \mathbf{1}_{c_i \in s}\right)^2$,
where $\bar{\boldsymbol{\phi}^r} = {\bar{\phi}^r_i,\text{ for }c_i\in\mathbf{S}^r}$. We apply the prior equations for normalization and surrogate Shapley
to these approximate Shapley values to get the surrogate approximate Shapley values, $\Phi^r_i$ for $c_i\in \mathcal{C}$. Hence, the global model accuracy $A^r$ in the optimization problem, Eqn.~\ref{eqn:ilp}, can be replaced as, $A^r = \sum_{c_i\in\mathcal{S}^r}\Phi^{r-1}_i$, for the candidate client cohorts.


\subsubsection{Efficient Cohort Evaluation Using Coreset Sampling.}
As the number of evaluations in exhaustive Shapley (ExSH) grows exponentially with the number of clients and in Kernel Shapley (KSH) scales linearly, our next objective is to make their evaluations faster. We use a \textit{coreset sampling} methodology based on the greedy facility location algorithm, inspired by the coreset construction~\cite{mirzasoleiman2020coresets}. Similar to CRAIG~\cite{mirzasoleiman2020coresets}, which uses sub-modular optimization to select representative training data subsets, we apply a facility location-based strategy to construct a representative subset of test data for efficient cohort evaluation.

Given an original test dataset \(\mathcal{D}_v\) with $n=|\mathcal{D}_v|$, an optional feature extractor, \(\varphi\), a subset size ratio \(\lambda \in (0.0,1.0)\), a minimum per-class sample threshold \(m_{\text{min}}\), and a cosine distance metric \(d\), the objective is to construct a coreset \(\mathcal{D}' \subset \mathcal{D}_v\) with approximately \(\lambda \cdot n\) samples while ensuring balanced class representation. The target sample count per class is, $m = \max\left(\left\lfloor \frac{N_{\text{target}}}{k} \right\rfloor, m_{\text{min}}\right)$, where 
\(k\) denotes the number of unique classes and
$N_{\text{target}} = \max\left(\lfloor \lambda \cdot n \rfloor, k \cdot m_{\text{min}}\right)$.

For each class \(y_j\), the subset \(\mathcal{D}_{v,j}\) is extracted and processed using the greedy facility location algorithm similar to~\cite{mirzasoleiman2020coresets}.
It first selects the \textit{medoid}, the point with the minimum sum of distances to all other points,$i_{\text{medoid}} = \arg\min_{i} \sum_{j} \Delta_{i,j}$,
where \(\Delta_{i,j} = d(x_i, x_j)\) represents the pairwise distance matrix.
Subsequent points are iteratively selected to maximize the \textit{marginal gain} in feature space coverage,$i_{\text{next}} = \arg\max_{i \notin \mathcal{R}} \sum_{j} \left(\mu_j - \min(\mu_j, \Delta_{i,j})\right)$ where \(\mathcal{R}\) is the set of selected samples, \(\mu_j = \min_{s \in \mathcal{R}} \Delta_{s,j}\) denotes the minimum distance from \(x_j\) to any selected sample \(s \in \mathcal{R}\), and \(\Delta_{i,j}\) represents the distance between sample \(x_i\) and \(x_j\).
This ensures that the selected points are well-distributed across the feature space. 

In \method, we use coreset sampling for cohort model evaluation during Shapley value computation, with the original dataset serving as the benchmark for final global model assessment. This involves constructing \(\mathcal{D}'\) using the coreset sampling algorithm, evaluating cohort models with \(\mathcal{D}'\) through ExSH or KSH methods, and then assessing the final aggregated global model on \(\mathcal{D}_v\).


\subsection{Energy-Time Pareto Estimation}

\label{sec:method:powertrain}

\begin{table}[t]

\centering
\scriptsize
\caption{Specifications of NVIDIA Jetson devices used in our experiments.}

\label{tbl:jetson_device_specs}
\begin{tabular}{l|C{2.3cm}|C{2.3cm}|C{2.3cm}|C{2.3cm}}
\hline
 & \textbf{Jetson NX Xavier} & \textbf{Jetson Orin Nano} & \textbf{Jetson AGX Xavier} & \textbf{Jetson Orin AGX} \\ \hline\hline
\textbf{CPU Arch./cores} & ARM Carmel/6c & ARM A78AE/6c & ARM Carmel/8c & ARM A78AE/12c\\ \hline
\textbf{CPU freq.} & 0.11--1.9GHz & 0.11--1.5GHz & 0.11--2.2GHz & 0.11--2.2GHz \\ \hline
\textbf{GPU Arch./cores} & Volta/384c & Ampere/1024c & Volta/512c & Ampere/2048c \\ \hline
\textbf{GPU freq.} & 0.11--1.1GHz & 0.30--0.62GHz & 0.11--1.3GHz & 0.11--1.3GHz \\ \hline
\textbf{Mem./Swap} & 8GB/4GB & 8GB/4GB & 32GB/16GB & 32GB/16GB \\ \hline
\textbf{Mem freq.} & 0.20--1.8GHz & 0.20--2.1GHz & 0.20--2.1GHz & 0.20--3.1GHz \\ \hline
\end{tabular}%

\end{table}

\method leverages our prior work~\cite{prashanthi2024powertrain} on a transfer-learning approach with limited profiling to predict power and time consumption for training a DNN using a given power mode on accelerated Jetson devices. We use \textit{reference Neural Network (NN) models} for power and time predictions that were earlier trained using exhaustive (one-time) profiling for training ResNet-18 DNN on ImageNet dataset for $4,368$ power modes on a Jetson Orin AGX. We need to generalize these prediction models to the three other edge device types (clients) we use (Table~\ref{tbl:jetson_device_specs}) and two other DNN models we train using FL (Table~\ref{tab:model-details}). Instead of performing costly profiling of 1000s of power modes for each device--DNN workload combination, we instead profile 90 randomly sampled power modes from each workload to measure their observed minibatch times and power load for one epoch of training.
This concise profiling data is used to re-train the reference NN models for power and time predictions~\cite{prashanthi2024powertrain}, and reduces profiling effort by $\approx 96\%$ from an estimated $1,075$ hours.


We use these transfer learned models to predict the power and training time for these workloads for \textit{all their power modes}. This is used to estimate the per-round training time and energy usage for each DNN on a device-type workload and shown as a scatter plot of energy vs. time. We identify the set of power modes that lie on the Pareto front of this scatter plot, i.e., the power mode gives the lowest training time for that energy budget. These give the possible set of viable power modes to solve \textit{ILP-PM}.
As Fig.~\ref{fig:pareto-errors} shows, we see low prediction errors of $\approx 7\%$ for power and $\approx 9\%$ on time for these power modes.


\section{Experiments and Results} 
\label{sec:results}

We perform extensive experiments to comparatively evaluate \method with simple and SOTA FL approaches, for federated learning of three DNN--dataset workloads using two accelerated edge clusters simulated with real-world traces.

\subsection{Experimental Setup}
\label{subsec: exp_setup}

\begin{table}[t]

  \centering
  \scriptsize
  \setlength{\tabcolsep}{0pt}
  \caption{Models, datasets and key FL hyperparameters.}
  \label{tab:model-details}
  \begin{tabular}{@{}%
      >{\raggedright\arraybackslash}p{1.7cm}|
      >{\centering\arraybackslash}p{1.4cm}
      >{\centering\arraybackslash}p{1.3cm}
      >{\raggedright\arraybackslash}p{1.7cm}
      >{\raggedright\arraybackslash}p{1.2cm}
      >{\centering\arraybackslash}p{1.1cm}
      >{\centering\arraybackslash}p{1.8cm}
      >{\raggedright\arraybackslash}p{2cm}
    @{}}
    \hline
    \textbf{Model} & \# \textbf{params} & \# \textbf{layers} & \textbf{Task} & \textbf{Dataset} & \textbf{Classes} & \textbf{Train/Test} & \textbf{Hyperparams} \\
    \hline\hline
    LSTM         & 0.166M & 3  & Language detection                          & Tatoeba       & 8   & 71.1k/1.6k & epochs=3 \\ \hline
    MobileNetV2  & 2.3M   & 53 & Image classification                        & CIFAR‑10      & 10  & 50k/10k   & dropout=0.2, epochs=3 \\ \hline
    ResNet‑18              & 11M    & 18 & Image classification& ImageNet Subset  & 200 & 200k/10k  & dropout=0.1, epochs=2 \\
    \hline
  \end{tabular}

\end{table}

\begin{table}[t]

\centering
\scriptsize
\caption{Non-IID datasets and heterogeneity metrics of clients for FL training.}
\label{tab:partition_characteristics}
\begin{tabular}{ll||c|rR{2.25cm}}
\hline
\textbf{Model} & \textbf{Dataset} & \textbf{Partitions} & \textbf{\#samples per part} & \textbf{Jensen-Shannon divergence} \\ \hline\hline
\multirow{2}{*}{LSTM} & \multirow{2}{*}{Tatoeba} & 12 & $\mu=5923$, $\sigma=2819$ & 0.31 \\ \cline{3-5} 
 &  & 48 & $\mu = 1480$, $\sigma = 1002$ & 0.34 \\ \hline
\multirow{2}{*}{MobileNetV2} & \multirow{2}{*}{CIFAR10} & 12 & $\mu=4167$, $\sigma= 1901$ & 0.44 \\ \cline{3-5} 
 &  & 48 & $\mu=1042$, $\sigma = 922$ & 0.43 \\ \hline
ResNet-18 & ImageNet Subset & 12 & $\mu=16,667$, $\sigma= 325$ & 0.34 \\ \hline
\end{tabular}%

\end{table}


\paragraph{Models and Datasets.} 
We run three diverse DNN workloads to comprehensively assess the performance of our proposed FL approach:
\textit{LSTM} for language detection the Tatoeba dataset, and image classification using \textit{MobileNetV2} trained on CIFAR10, and the significantly larger \textit{ResNet-18} trained on a subset of ImageNet with 200 classes.
The training hyper\-parameters are provided in Table~\ref{tab:model-details}. $lr=1\times10^{-4}$ and $bs=16$ for all. These models have varying complexity~(0.166--11M parameters) and dataset domains~(text and image). Each dataset was split in a non-IID manner into 12 and 48 partitions, one per client in our two FL configurations~(Tab.\ref{tab:partition_characteristics}). 
Tatoeba and ImageNet data are partitioned 
with each class divided into $\left\lceil\frac{c\times\delta}{l} \right\rceil$ shards, where $c$ is the number of clients, $\delta$ the labels per client, and $l$ the total classes in the dataset. We assigned $\delta=3$ labels per client for Tatoeba with 12 partitions, $\delta=2$ for 48 partitions, and $\delta=60$ for ImageNet. CIFAR10 is split using a Dirichlet-based partitioning with concentration parameter $\alpha=0.05$ for 12 and 48 clients. This is a more challenging non-IID setup with more diverse class distributions across clients. 


\paragraph{Simulation Setup.}
We host a FL simulation environment on a GPU workstation with AMD Ryzen 9 7900X 12-Core CPU@5.73GHz, 128GB RAM, and a NVIDIA GeForce RTX 4090 GPU with 24GB. For the 12-device cluster, we model heterogeneous edge accelerators with 2 Jetson Orin AGX, 2 Jetson AGX Xavier, 4 Jetson Xavier NX, and 4 Jetson Orin Nanos (Table~\ref{tbl:jetson_device_specs}). For the 48-device setup, we have 12 of each device type to assess scalability. While our FL model training and convergence properties are based on simulating FL execution, the energy and runtime values reported are based predictions from PowerTrain for each DNN workload on the corresponding Jetson edge accelerators.


\paragraph{Baseline Strategies.} 

We evaluate two variants of \method, with exhaustive Shapley values (\textbf{FedJ-ex}) and Kernel Shapley values (\textbf{FedJ-k}) -- the former is more accurate but slower while the latter is faster. We also compare with several synchronous FL strategies. As a baseline, we run FedAvg~\cite{McMahan2016CommunicationEfficientLO} with $\gamma$ clients randomly chosen in each round (\textbf{RND}). We use a simpler Shapley-based strategy with probabilistic sampling (\textbf{ExSH}). In the first round, the Shapley value for every client is set to 1, leading to $\gamma$ random clients being sampled. On receiving the local models, the server updates the sampling weights of these clients with the surrogate exhaustive Shapley values, $\tilde{\phi}_i^r$.
The server then samples a set of clients for the next round using these updated weights.
We also adapt this same strategy but with the Kernel Shapley approximations (\textbf{KSH}). These two show the benefits of our Shapley optimizations into \method, which go beyond the default Shapley-based FL approaches.
Lastly, we compare against a state-of-the-art energy-aware FL strategy, Energy Saving Client Selection~(\textbf{ESCS})~\cite{escs}.
This selects clients based on multiple criteria, including battery level, training time capacity, and network quality based on a client utility score. We modified this to use the \textit{system-based utility} and consider only the terms containing loss and the time, and use their deterministic client selection strategy. The training loss for future rounds is obtained by \textit{training all clients} in the first round -- a deviation from our problem definition.


\subsection{Evaluation of Strategies}
\label{subsec: sim_experiments}

\begin{figure}[t]

\centering
\subfloat[LSTM 6/12 \textit{(100MJ)}]{\label{subfig:lstm-6-12}
    \includegraphics[width=0.33\textwidth]{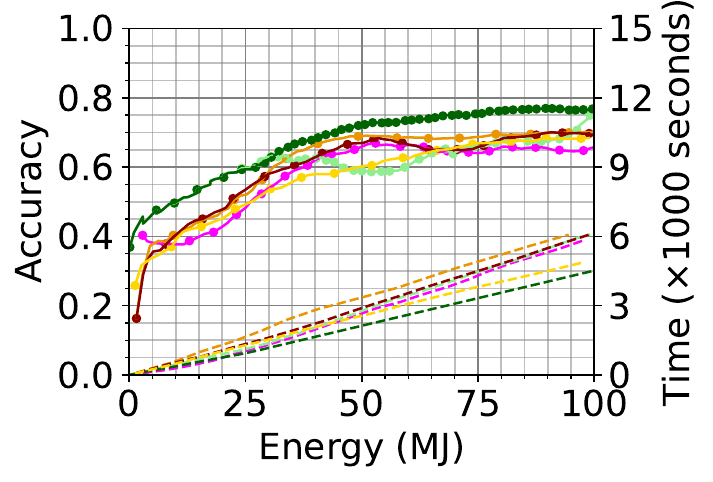}}
\subfloat[MobileNet 6/12 \textit{(200MJ)}]{\label{subfig:mob-6-12}
    \includegraphics[width=0.33\textwidth]{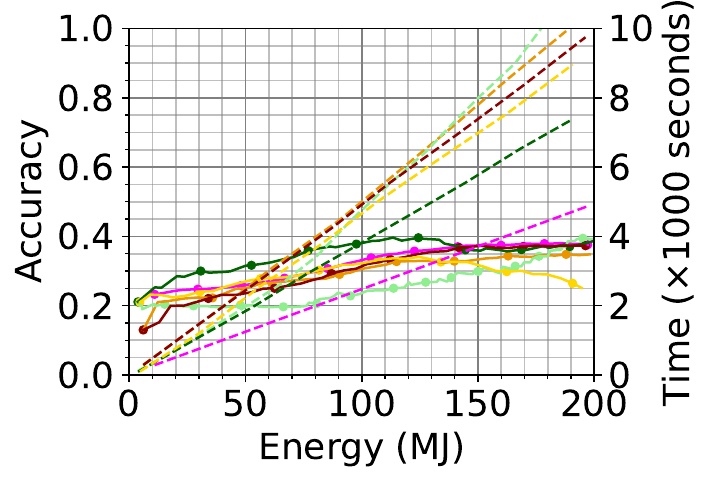}}
\subfloat[ResNet 6/12 \textit{(600MJ)}]{\label{subfig:res-6-12}
    \includegraphics[width=0.33\textwidth]{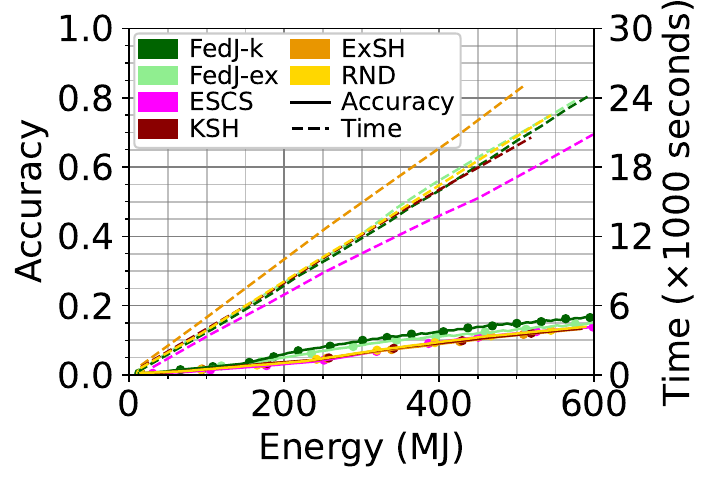}}\\

\caption{Accuracy achieved (left Y axis) and FL training time (right Y axis) as the global energy consumed increase (X axis) as FL rounds progress.}
\label{fig:acc-6/12}

\end{figure}

\begin{figure}[t]
    \centering
    
    \includegraphics[width=0.85\linewidth]{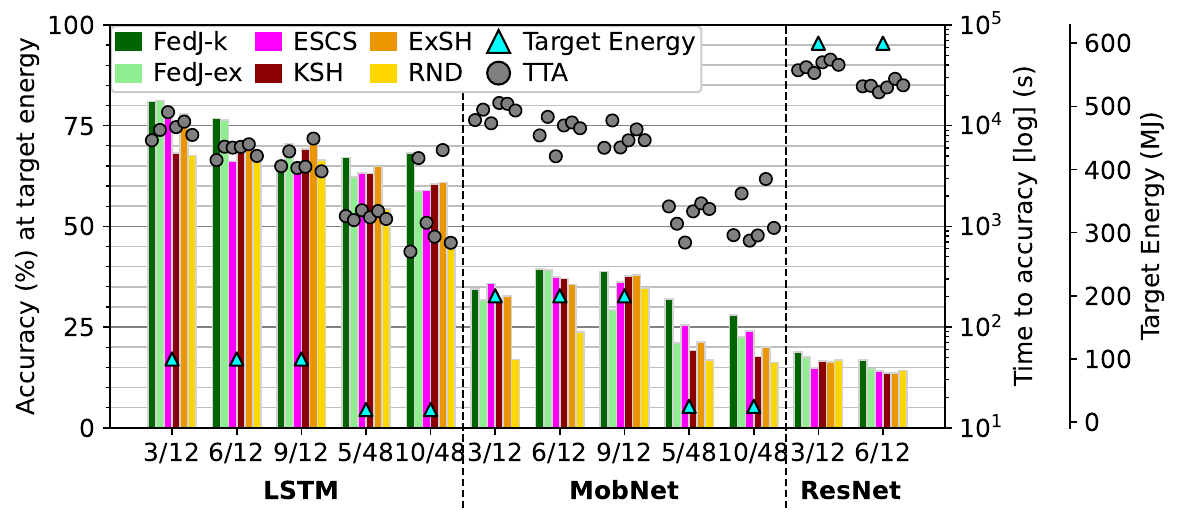}
    
    \caption{Accuracy at target energy--100MJ for LSTM-12, 20MJ for LSTM-48, 
    200MJ for MobNet-12, 15MJ for MobNet-48 and 500MJ for ResNet-18.}
    \label{fig:summary-100}

\end{figure}

We perform a detailed comparative analysis of FL training of the 3 workloads on the two clusters using our \ilpks and \ilpes, and against \random, \es, \ks and \escs. We also vary the \textit{client cohort sizes per round} as 3/12, 6/12, 9/12, 5/48 and 10/48, where $m/n$ means $m=\gamma$ clients are selected from $n$ clients. Since ResNet training is much slower, we only run it on 3/12 and 6/12, which takes 3h and 8h per experiment respectively.
Further, we set the \textit{global energy budget } for each configuration partly based on the DNN model sizes since larger models take longer to train and consume more energy. For example, LSTM takes $\approx0.2$MJ for one client to train one round on a 12-client cluster, while ResNet takes about $\approx2$MJ. Further, if the cluster has more clients, the data will have been partitioned such that each client has fewer training samples resulting in the energy usage per training round being lower per client.
Based on this, we set the energy budget ($\widehat{E}$) for 12-client/48-client setup as 100MJ/20MJ for LSTM, 200MJ/25MJ for MobNet and 600MJ/- for ResNet. 
\begin{figure}[t]

    \centering
    \includegraphics[width=0.85\linewidth]{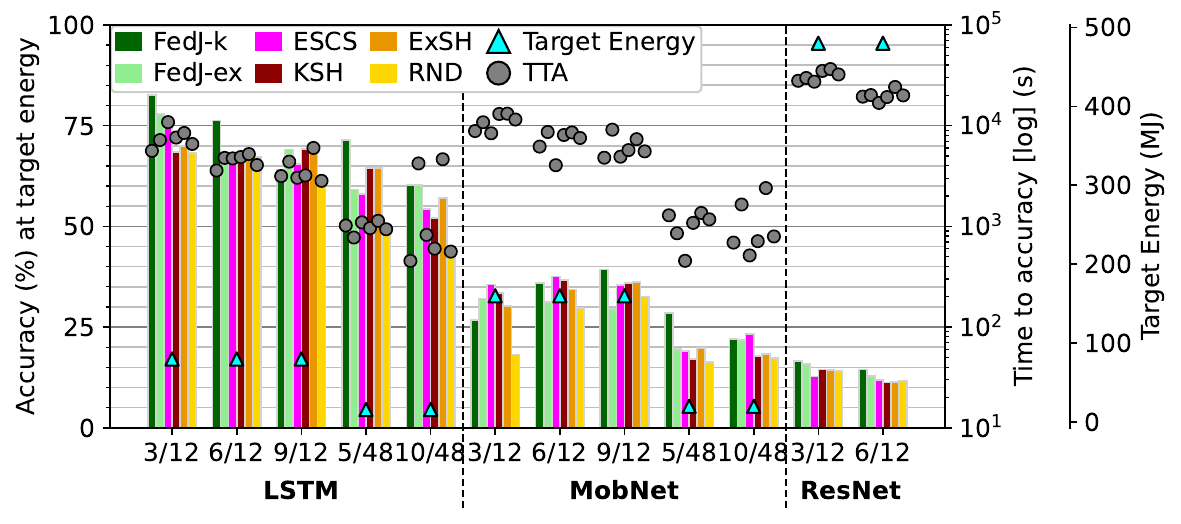}
    
    \caption{Accuracy at target energy 80\% of original--80MJ for LSTM-12, 16MJ for LSTM-48 , 160MJ for MobNet-12, 20MJ for MobNet-48, 480MJ for Resnet-12}
    \label{fig:summary-80}
    
\end{figure}


\paragraph{Relaxed Energy Budget.}
We summarize the \textit{FL training accuracy} based on energy budgets and \textit{training time} in Fig.~\ref{fig:summary-100} for the different setups and strategies.  The target energy budgets were set based on the minimum aggregate energy required for each model to reach a stable accuracy in preliminary experiments by the most energy-efficient strategy for the configuration. E.g., LSTM-12 and MobNet-12 require at least 100~MJ and 200~MJ, respectively, to converge on accuracy, while the more efficient LSTM-48 and MobNet-48 reach similar plateaus at 20~MJ and 15~MJ. ResNet-18, being deeper and requiring more numerical operations, only stabilized in performance above 500~MJ. These thresholds were thus selected to ensure each model is evaluated at its optimal energy–accuracy tradeoff point, avoiding both under-fitting and wasted energy. For a subset, 6/12, Fig.~\ref{fig:acc-6/12} also shows the increase in global model accuracy (left Y axis) and time taken (right Y axis) over the rounds as the energy budget is used up (X axis).

We see that except LSTM-9/12 and MobNet-3/12, \ilpks outperforms all other strategies on the highest accuracy achieved within the energy budget. This is because \ilpks usually performs the most number of model updates within the energy budget~(e.g., 49\% more than \random and \es on average across cohort sizes), which facilitates \ilpks's lower per-round energy usage compared to the other baselines due to the partial participation of clients. This is even more evident for larger cohorts, where \ilpks has a per-round energy usage of 4.65MJ, 2.56MJ and 0.71MJ  for MobNet 6/12, 9/12 and 10/48, respectively as compared to \random's 5.4MJ, 8MJ and 2.2MJ. We note from Fig.~\ref{fig:acc-6/12} that despite picking fewer clients per round \ilpks also achieves faster convergence compared to the baseline strategies, thanks to the Shapley value based client contribution measurement. While our approach requires an initial energy investment of 400-8000J for the profiling phase, it forms a very tiny fraction of the energy for the FL session and is significantly offset by energy savings of approximately 50-400MJ per workload. We note that the FL server itself can run on the cloud and will not add to the overall energy cost of the devices.

Usually, we expect the accuracy of strategies to increase as client participation increases, which we do not see in Fig.~\ref{fig:summary-100}. This is expected in an energy-constrained case. However, as participation increases, the energy cost per round also increases. E.g., in ResNet, the average energy consumed per round for \ks increases from 8MJ to 17MJ from 3/12 to 6/12, leading to the strategies having an average of $\approx$50\% fewer rounds performed. This trend, however, does not hold for MobileNet, where, due to the extreme skew of the data, the global model benefits from increased participation. As expected, \ilpes performs on par with \ilpks on smaller cohort sizes, reaching highest or second highest accuracies in two out of the three 3/12 configurations. But we note the decreases in accuracy at target energy as its round-time increases 94\% for LSTM and 73\% for MobNet for cohort sizes of 5 with respect to 10. This highlights the exponential time complexity of model evaluation for exhaustive Shapley value strategies and the benefits of \ilpks.


\paragraph{Restricted Energy Budget.}
Now, we reduce the global energy budget by 20\% over the previous and observe the behaviour of these FL strategies for a more sustainable and lower carbon-footprint target. 
We see the same convergence trends hold in Fig.~\ref{fig:summary-80}, which show the accuracies and runtimes achieved when 80\% of the earlier energy budget was reached for each model-configuration pair. From Figs.~\ref{fig:summary-100} and \ref{fig:summary-80}, we see that \ilpks and \ilpes gain the highest increment in accuracy across all strategies in when the energy budget is relaxed by 20\%, further highlighting their rapid convergence.

\begin{figure*}[t]
    \centering
    \begin{minipage}[b]{0.49\textwidth}%
        \centering%
        
        \includegraphics[width=\textwidth]{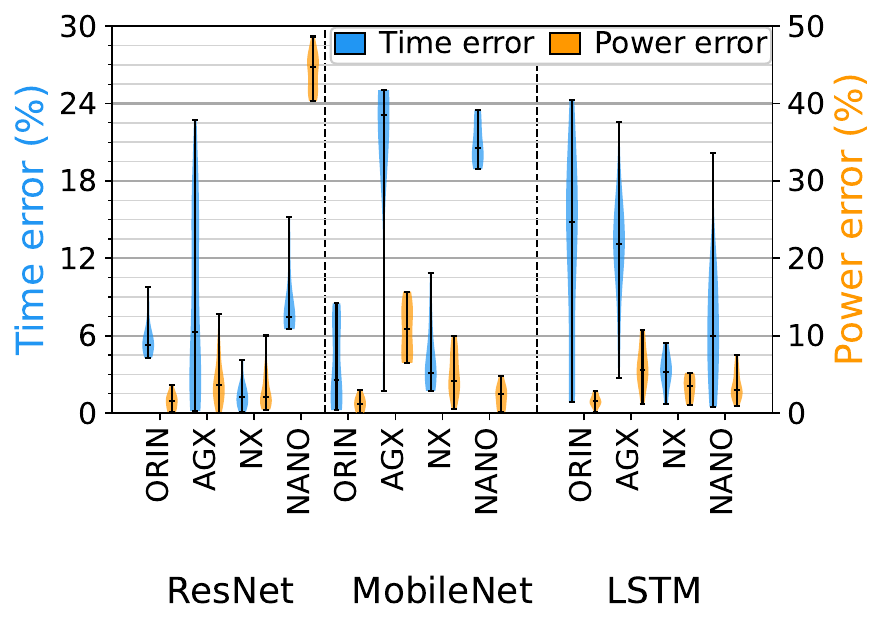}
        
        \caption{Time and power prediction errors for the Energy-Time Pareto points.}%
        \label{fig:pareto-errors}%
    \end{minipage}%
    \quad
    \begin{minipage}[b]{0.45\textwidth}%
        \centering%
        
        \includegraphics[width=\textwidth]{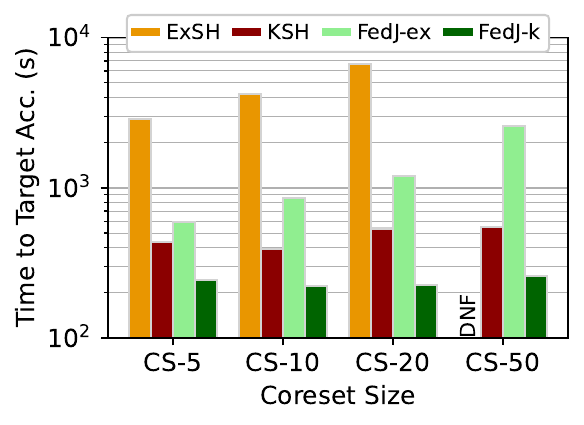}
        
        \caption{Time-to-accuracy (16.8\%) for \es, \ks, \ilpes and \ilpks for MobNet-10/48 for Coreset sampling 5\%, 10\%, 20\%, and 50\%.}%
        \label{fig:coreset-tta}
    \end{minipage}%

\end{figure*}


\paragraph{Effect of Coreset Sampling Size.}

\label{subsec: coreset_eval}

The size of the Coreset-sampled evaluation data for strategies based on Shapley values represents a trade-off between lower computation times~(smaller datasets will have faster cohort evaluations) and better evaluation of client contributions. Here, we explore this by evaluating the performance of \es, \ks, \ilpes and \ilpks in the MobNet-12/48 setting for four different Coreset sampling percentages: 5\%~(CS-5), 10\%~(CS-10), 20\%~(CS-20) and 50\%~(CS-50). Fig.~\ref{fig:coreset-tta} summarizes the \textit{time-to-target accuracy (TTA)} for CS-5, CS-10, CS-20 and CS-50 for \es, \ks, \ilpes and \ilpks. The sensitivity of a strategy to Coreset size is directly linked to the number of model evaluations it performs per round, creating significant performance variations as the validation data constraints change.
As expected, \es and \ilpes are the most affected by the size of the validation data since for MobNet-10/48, these strategies perform $1,023$ model evaluations per round. \es takes $4,000$ secs to reach target accuracy for CS-10, which increases to $6,680$~secs for CS-20. Similarly for \ilpes, $859$ secs are needed to reach target accuracy for CS-10, which increases to $1,206$ secs for CS-20 and $2,568$ seconds for CS-50. \ks on the other hand, performs only 30 evaluations per round, and thus its runtime increases only slightly from $436$~secs for CS-10 to $390$ and $535$~secs for 20 and 50. \ilpks does a maximum of 30 evaluations per round, leading it to have the fastest runtimes of $240$, $220$, $225$ and $250$~secs across the four Coreset sizes.

With the smallest Coreset size of CS-5, all strategies see an increase in TTA, as compared to CS-10. This is because after a point, the dataset is too small to quantify client contributions. With the smallest size of CS-5, all strategies experience an increase in TTA compared to CS-10. This trend is attributed to the dataset becoming too small to accurately quantify client contributions, thereby affecting model evaluations and utility estimation. ML literature emphasizes that the quality and size of the validation set plays a critical role in model testing. A smaller validation set can lead to noisy utility estimates, reducing the model's ability to gauge client contributions effectively. Thus, it is important to balance validation data size with computational constraints.


\section{Conclusions}
\label{sec:conclude}

We address the problem of limiting global energy budgets for FL using heterogeneous Jetson edge accelerators over non-IID data. Our unique bi-level ILP formulation leverages approximate Shapley values and energy-time prediction models to efficiently solve this problem. Our \method framework achieves superior training accuracies compared to SOTA and baselines for diverse energy budgets and realistic experiment configurations. In future work, we plan to run more extensive experiments on physical edge clusters and complement the ILP-based approach with a faster probabilistic heuristic that continues to balance per-round times and global model quality. We aim to extend our approach to more heterogeneous platforms, including mobile devices and accelerators like TPUs and FPGAs. We further plan on performing in-depth real-world validation to assess the impact of network conditions, client dropouts and map energy usage to CO$_2$ emissions to better assess the environmental impact of our methodology.

{\fontsize{9pt}{9pt}\selectfont
\subsubsection{Disclosure of Interest.~} The authors declare that they have no known competing financial interests or personal relationships that could have appeared to influence the work reported in this paper.
}

\bibliographystyle{plain}
\bibliography{references}

\begin{thebibliography}{10}

\bibitem{chen2023eefl}
Rui Chen et~al.
\newblock Eefl: High-speed wireless communications inspired energy efficient federated learning over mobile devices.
\newblock In {\em MobiSys}, 2023.

\bibitem{cho2022towards}
Yae~Jee Cho, Jianyu Wang, and Gauri Joshi.
\newblock Towards understanding biased client selection in federated learning.
\newblock In {\em AISTATS}, 2022.

\bibitem{data-diversity}
Kevin Hsieh, Amar Phanishayee, Onur Mutlu, and Phillip Gibbons.
\newblock The non-{IID} data quagmire of decentralized machine learning.
\newblock In {\em ICML}, 2020.

\bibitem{jia2019towards}
Ruoxi Jia et~al.
\newblock Towards efficient data valuation based on the shapley value.
\newblock In {\em AISTATS}, 2019.

\bibitem{sys-diversity}
Gyudong Kim et~al.
\newblock Heteroswitch: Characterizing and taming system-induced data heterogeneity in federated learning.
\newblock In {\em MLSys}, 2024.

\bibitem{fedprox}
Tian Li et~al.
\newblock Federated optimization in heterogeneous networks.
\newblock {\em MLSys}, 2020.

\bibitem{liu2023green}
Jingchu Liu et~al.
\newblock Green federated learning: A sustainable framework for energy-constrained iot networks.
\newblock {\em IEEE IoT J.}, 2023.

\bibitem{lundberg_2017_kernelshap}
Scott~M. Lundberg and Su-In Lee.
\newblock A unified approach to interpreting model predictions.
\newblock In {\em NeurIPS}, 2017.

\bibitem{escs}
Filipe Maciel et~al.
\newblock Federated learning energy saving through client selection.
\newblock {\em Pervasive Mob. Comput.}, 2024.

\bibitem{McMahan2016CommunicationEfficientLO}
H.~B. McMahan et~al.
\newblock Communication-efficient learning of deep networks from decentralized data.
\newblock In {\em AISTATS}, 2016.

\bibitem{mirzasoleiman2020coresets}
Baharan Mirzasoleiman, Jeff~A. Bilmes, and Jure Leskovec.
\newblock Coresets for data-efficient training of machine learning models.
\newblock In {\em ICML}, 2020.

\bibitem{sigmetrics-prashanthi}
S.K. Prashanthi et~al.
\newblock Characterizing the performance of accelerated jetson edge devices for training deep learning models.
\newblock {\em POMACS}, 2022.

\bibitem{prashanthi2024powertrain}
S.K. Prashanthi et~al.
\newblock Powertrain: Fast, generalizable time and power prediction models to optimize dnn training on accelerated edges.
\newblock {\em FGCS}, 2024.

\bibitem{sum_2023_shapleyFL}
Qiheng Sun et~al.
\newblock Shapleyfl: Robust federated learning based on shapley value.
\newblock In {\em KDD}, 2023.

\bibitem{wang2020principled}
Hongyi Wang et~al.
\newblock Principled weight sharing in federated learning.
\newblock In {\em ICLR}, 2020.

\bibitem{Yu2022energyaware}
Chong Yu, Shuaiqi Shen, Kuan Zhang, Hai Zhao, and Yeyin Shi.
\newblock Energy-aware device scheduling for joint federated learning in edge-assisted internet of agriculture things.
\newblock In {\em WCNC}, 2022.

\bibitem{zhang2021gtg}
Zelei Zhang et~al.
\newblock Gtg-shapley: Efficient and accurate participant contribution evaluation in federated learning.
\newblock In {\em TPDS}, 2021.

\end{thebibliography}

\newpage
\appendix
\section*{Appendix}
\section{Scalability of \ilpks}
\begin{figure}
    \centering
    \includegraphics[width=\linewidth]{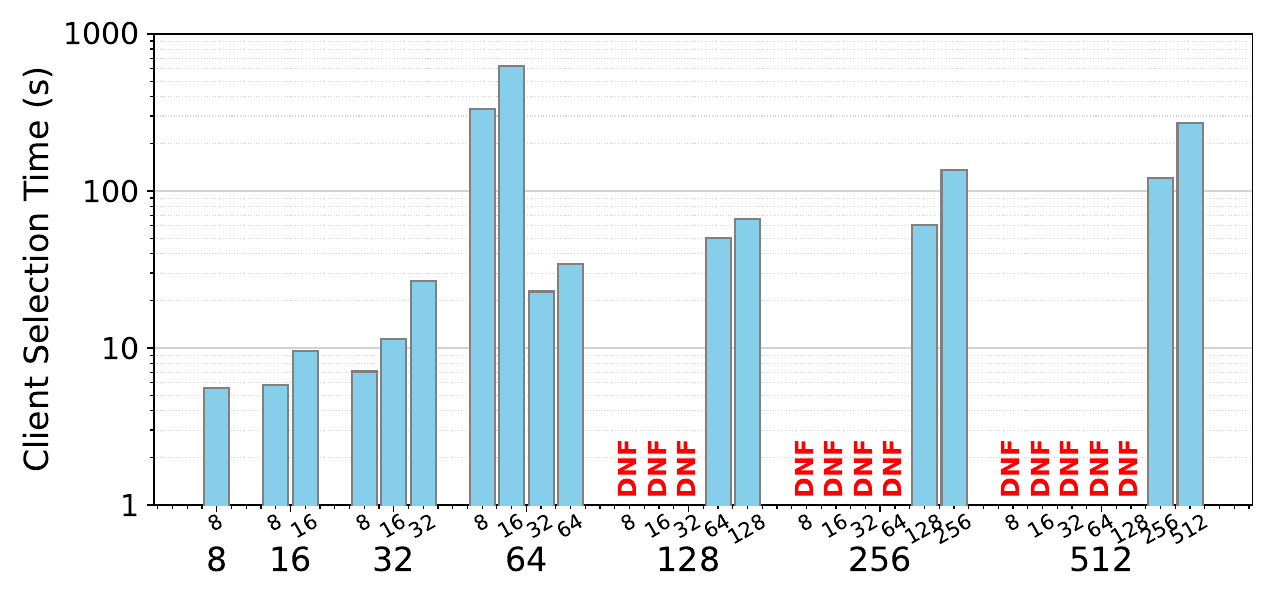}
    \caption{Client selection time for different client cohort sizes and different total number of clients available. The inner x-axis shows the cohort size and the outer x-axis shows the system size. The red star depicts client selection did not finish in 1hour. All times are averaged over 5 rounds.}
    \label{fig:scalability}
\end{figure}

The scalability analysis of \ilpks, as shown in the Fig. \ref{fig:scalability}, illustrates how client selection time varies with both the number of clients selected per round (inner ticks) and the total number of available clients (outer ticks). The results indicate that as the total client pool increases from 8 to 512, the client selection time for \ilpks grows rapidly since the search space of the ILP grows rapidly. Even as the number of clients selected per round rises (from 8 up to 256), the increase in selection time remains moderate, highlighting the efficiency of the algorithm ill-suited for large-scale federated learning deployments, and needs to be replaced by a heuristic based method.

\section{Energy to Acccuracy}
\begin{figure}
    \centering
    \includegraphics[width=0.9\linewidth]{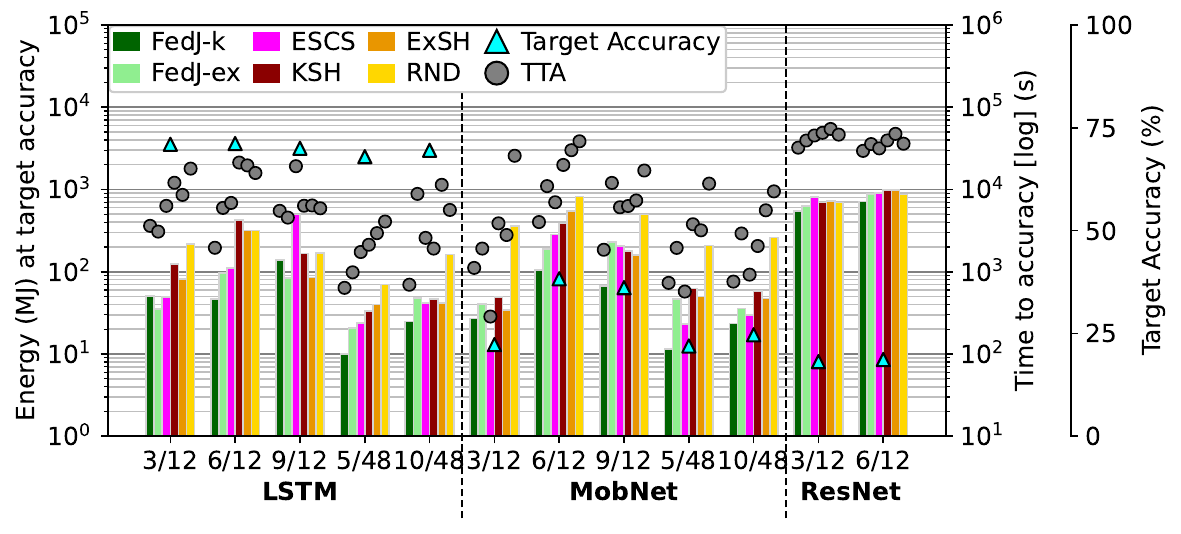}
    \caption{Energy and time-to-accuracy tradeoffs for federated learning methods (\ilpks, \ilpes, \escs, \ks, \es, \random) and deep learning baselines, showing specialized federated approaches achieve higher accuracy with significantly lower energy and faster convergence.}
    \label{fig:eta}
\end{figure}

Fig.\ref{fig:eta} demonstrates critical performance tradeoffs between energy consumption and model accuracy across various federated learning algorithms. Traditional deep learning architectures (\textbf{LSTM} ,  \textbf{MobNet} , and \text{ResNet}) exhibit significantly higher energy requirements (approaching 1200 MJ) compared to federated optimization techniques like \ilpks and  \ilpes at equivalent accuracy targets. The logarithmic time-to-accuracy plot reveals convergence times spanning $10^1$ to $10^6$  seconds, with specialized approaches \escs and \ks achieving faster convergence at lower energy costs. At higher accuracy levels (75–100\%), energy consumption increases nonlinearly for most methods, particularly impacting resource-intensive architectures like \textbf{ResNet}. While \es demonstrates superior energy efficiency, this optimization comes at the cost of reduced maximum achievable accuracy compared to \ilpes. The \random baseline exhibits the worst energy-time-accuracy tradeoffs, emphasizing the importance of structured optimization in federated environments. These results highlight the need for algorithm selection based on deployment constraints-energy-constrained edge devices may prioritize \es or \ilpks, while latency-tolerant applications could leverage \ilpes for higher final accuracy.

\end{document}